\documentclass[lettersize,journal]{IEEEtran}

\usepackage{amsmath}
\usepackage{enumitem}
\usepackage{lineno} 
\usepackage{hyperref}
\usepackage{array}
\usepackage{cite}

\usepackage{pgfplots}
\pgfplotsset{compat=1.8}
\usepackage{graphicx}
\usepackage{multirow}
\usepackage{verbatim} 
\usepackage[utf8]{inputenc}
\usepackage{csquotes}
\usepackage{booktabs}
\usepackage{algpseudocode}
\usepackage{algorithm}
\usepackage{tikz}
\usetikzlibrary{shapes.geometric, arrows}
\usepackage{xcolor}
\usepackage{float}
\usepackage{colortbl}
\usepackage{threeparttable}
\usepackage{amssymb}

\def\BibTeX{{\rm B\kern-.05em{\sc i\kern-.025em b}\kern-.08em
    T\kern-.1667em\lower.7ex\hbox{E}\kern-.125emX}}

\begin{document}

\title{Embracing the Generative AI Revolution: Advancing Tertiary Education in Cybersecurity with GPT}

\author{Raza Nowrozy, and David Jam
\thanks{Manuscript submitted \today. \textit{(Corresponding author: Raza Nowrozy)}}
\thanks{Raza Norwozy is with Victoria University, Melbourne, VIC, Australia (email: raza.nowrozy@live.vu.edu.au). David Jam is with Vertical Scope, Melbourne, VIC, Australia (email: d.jam@verticalscope.com.au). }
}

\maketitle

\begin{abstract}
	The rapid advancement of generative \textit{Artificial Intelligence} (AI) technologies, particularly \textit{Generative Pre-trained Transformer} (GPT) models such as ChatGPT, has the potential to significantly impact cybersecurity. In this study, we investigated the impact of GPTs, specifically ChatGPT, on tertiary education in cybersecurity, and provided recommendations for universities to adapt their curricula to meet the evolving needs of the industry. Our research highlighted the importance of understanding the alignment between GPT's ``mental model'' and human cognition, as well as the enhancement of GPT capabilities to human skills based on Bloom's taxonomy. By analyzing current educational practices and the alignment of curricula with industry requirements, we concluded that universities providing practical degrees like cybersecurity should align closely with industry demand and embrace the inevitable generative AI revolution, while applying stringent ethics oversight to safeguard responsible GPT usage. We proposed a set of recommendations focused on updating university curricula, promoting agility within universities, fostering collaboration between academia, industry, and policymakers, and evaluating and assessing educational outcomes.
\end{abstract}

\begin{IEEEkeywords}
Cybersecurity; ChatGPT; Generative Pre-trained Transformer; University degree; Generative AI
\end{IEEEkeywords}

\section{Introduction}
\label{sec:introduction}
\IEEEPARstart{T}{he} rapid advancement of generative \textit{Artificial Intelligence} (AI) has resulted in significant changes to various industries, including cybersecurity \cite{dash2023chatgpt,mcintosh2023culturally,haleem2022era,mijwil2023chatgpt,truong2020artificial}. \textit{Generative Pre-trained Transformers} (GPTs), such as OpenAI's ChatGPT\footnote{https://openai.com/blog/chatgpt}, are a specific type of \textit{Large Language Models} (LLMs) that have emerged as a driving force in the AI revolution. With advanced \textit{Natural Language Understanding} (NLU) capabilities like sentiment analysis, controlled text generation, and implicit knowledge representation, GPTs demonstrate the potential to automate many tasks that used to be considered exclusive to human beings \cite{bubeck2023sparks,dash2023chatgpt,mcintosh2023google,eloundou2023gpts,haleem2022era}. In the cybersecurity domain, generative AI can be employed to formulate cybersecurity \textit{Governance, Risk, and Compliance} (GRC) policies \cite{mcintosh2023harnessing,nowrozy2024gpt,mcintosh2024cobit}, bolster cybersecurity defenses \cite{taofeek2022cognitive}, and even create malware or cyber threats \cite{liebowitz2021deception,mansfield2023weaponising,ranade2021generating}. This use underscores the importance of tertiary education providers incorporating generative AI into cybersecurity degree curricula. Tertiary institutions are often expected to play a crucial role in shaping the workforce, by providing graduates with the necessary knowledge and skills to excel in their chosen careers, and serving as a bridge between academic knowledge and practical application, to help students transition from the classroom to the workplace \cite{mishkind2014overview,tomlinson2017forms}. However, with the increasing integration of GPTs into industries, concerns have been raised about potential job losses and skill obsolescence, underlining the urgency for tertiary education providers to adapt their curricula to remain relevant and to prepare their students for the changing industry landscape, including the responsible and innovative use of generative AI in cybersecurity \cite{dwivedi2023so,mcguinness2021skills}.

Despite their best attempts, tertiary education institutions have sometimes been criticized for their slow response to the generative AI revolution, which may leave graduates unprepared for the changes and challenges faced in their careers \cite{bardhan2013responsive,bozkurt2023speculative,chatterjee2020adoption,popenici2017exploring}. In spite of the many disciplines offered in tertiary education, this study has chosen to focus on cybersecurity as the primary discipline of interest, because the cybersecurity industry is both highly sensitive to the adoption of AI, and has already demonstrated significant adoption of AI technologies, such as heuristic malware and threat detection, in their operations \cite{bertino2021ai,koroniotis2020holistic}. In light of this, tertiary institutions should assess the compatibility of their current cybersecurity curricula with the rise of GPTs, and proactively evaluate and update their curricula to address such changes, to ensure that their graduates are equipped with the most relevant and up-to-date knowledge and skills to remain competitive in the job market \cite{bozkurt2023speculative,chatterjee2020adoption,rudolph2023chatgpt}. This study aims to provide a comprehensive analysis of current topics and teaching approaches in cybersecurity tertiary education, with a focus on how well they align with industry expectations, the rise of GPTs, and the implications of its integration into the cybersecurity industries. By identifying potential gaps and areas for improvement, this research can contribute to the ongoing conversation around generative AI in education and help inform future curriculum development efforts.

The primary research problem is to evaluate the influence of GPT on tertiary education in cybersecurity, focusing specifically on universities. ChatGPT, a commercial implementation of GPT, is tailored for generating contextually relevant conversational responses, while sharing the core architecture with other GPT models. However, in this study, we focus on the common features shared among all GPT models. Our goal is to identify how universities can adapt their curricula to address potential job losses and skill obsolescence. The objectives of this study are: to assess the compatibility of current topics and teaching approaches in tertiary education with the new era of GPT, to evaluate the agility and responsiveness of universities to the evolving landscape of AI technologies, to identify the essential knowledge and skills for future graduates in response to the increasing use of GPT models in cybersecurity, and to provide recommendations for updating university curricula and fostering collaboration between academia, industry, and policymakers.

The major contributions of this study are listed as follows:
\begin{enumerate}[label=\arabic*)]
	\item We have extended the understanding of the influence of generative AI technologies, specifically GPT, on tertiary education in cybersecurity, and provided insights on how universities can adapt their cybersecurity curricula to meet industry needs.
	\item We have explored the alignment between GPT's ``mental model'' and human cognition, and enhancing GPT capabilities to human skills based on Bloom's taxonomy, which adds a unique perspective to the existing literature.
	
	\item We have proposed a set of comprehensive recommendations for updating university curricula, fostering agility within tertiary institutions, strengthening collaboration between academia, industry, and policymakers, and evaluating educational outcomes.
	
	\item We have contributed to the development of more holistic, relevant, and future-oriented cybersecurity education programs that effectively balance AI and human skills, ultimately preparing graduates to thrive in the rapidly evolving cybersecurity landscape.
\end{enumerate}

The rest of the article is organized as follows: Section \ref{sec:LiteratureReview} is a literature review of the related works. Section \ref{sec:TheoreticalFramework} lays out the theoretical framework used in this research. Section \ref{sec:Methodology} presents the methodology in acquiring and analyzing data. Section \ref{sec:Findings} unveils the findings of the study. Section \ref{sec:Discussion} discusses the findings of the study. Section \ref{sec:Recommendations} offers the recommendations to the tertiary institutions and stakeholders. Section \ref{sec:Conclusion} concludes the study, discusses its limitations, and suggests future research directions.\par

\section{Literature Review}
\label{sec:LiteratureReview}
In this section, we perform a literature review of the related works.

\subsection{AI and GPT models in cybersecurity}
The integration of generative AI technologies into cybersecurity has significantly transformed the industry, offering innovative solutions for detecting, analyzing, and mitigating security threats \cite{choi2021easy,yan2019automatically,yang2022generative,yinka2020review}. GPTs such as ChatGPT have been at the forefront of this transformation, providing unique capabilities in automating tasks like threat intelligence analysis, vulnerability identification, and intrusion detection \cite{choi2021easy,nam2021intrusion,yang2022generative}. Researchers highlighted the potential of GPTs to reduce the human workload and improve the overall efficiency of cybersecurity operations \cite{jun2021artificial,mijwil2023chatgpt}. Despite the advantages, concerns were raised about the potential misuse of GPT models for malicious purposes, such as generating phishing emails or crafting social engineering attacks \cite{karanjai2022targeted}. The literature on AI and GPT models in cybersecurity primarily focused on the benefits and potential risks, emphasizing the need for continuous development of AI-based security solutions, while also addressing ethical concerns and potential misuse \cite{lyn2019opportunities,truong2020artificial,vahakainu2019artificial}.

\subsection{Impact of GPT on tertiary education}
The increasing integration of LLMs into various industries prompted a growing interest in understanding their impact on tertiary education \cite{eloundou2023gpts}. LLMs, including GPT models, could enhance teaching and learning experiences through personalized learning, adaptive assessment, and intelligent tutoring systems \cite{bubeck2023sparks,eloundou2023gpts}. Additionally, GPT-driven educational tools were found to possess the potential to support students in developing critical skills such as problem-solving, creativity, and adaptability \cite{li2021natural,nori2023capabilities}. However, they also raised concerns about the potential displacement of human labor, leading to job losses and skill obsolescence \cite{dwivedi2023so,eloundou2023gpts,mcguinness2021skills}. The studies highlighted the need for tertiary education institutions to adapt their curricula to remain relevant and prepare students for the changing job market \cite{speight2013contested}.

\subsection{University responsiveness and policy statements}
The responsiveness of tertiary education institutions to the evolving landscape of AI technologies was a topic of growing importance in the literature, when universities had been criticized for their slow adaptation to technological advancements and the changing needs of industries \cite{ali2021bibliographical,bardhan2013responsive,bozkurt2023speculative,chatterjee2020adoption,junyent2008education,popenici2017exploring}. Some researchers argued that universities should be more agile in updating their curricula and incorporating emerging technologies like GPT models \cite{thurzo2023impact}. The \textit{United Nations Educational, Scientific and Cultural Organization} (UNESCO) and the \textit{Organisation for Economic Co-operation and Development} (OECD) both highlighted the potential benefits of implementing policy statements that outline universities' stance and approach to AI technologies, guiding curriculum development and fostering collaboration between academia, industry, and policymakers \cite{miao2021ai,organisation2018future}. A few examples of tertiary education institutions adapting their curricula and offering specialized courses or micro-accreditations related to AI technologies, including GPT, had been identified\footnote{https://news.it.ufl.edu/education/earn-ai-micro-credential-through-uf/}\textsuperscript{,}\footnote{https://powered.athabascau.ca/product?catalog=AI-Micro-Credential}. However, such examples remained limited, suggesting that more comprehensive efforts will be needed to ensure that tertiary education institutions are adequately preparing students for the future.

\subsection{Gaps in the existing research}
The extensive literature on AI technologies in tertiary education highlighted numerous advancements, yet several gaps remained. Firstly, there has been a scarcity of research examining the direct impact of GPT models, such as ChatGPT, on tertiary education and curricula. Most existing studies focused on AI technologies in general, primarily concentrating on earlier implementations like machine learning and data mining. Those existing studies looked into aspects such as personalized learning, adaptive learning systems, and automated assessment tools, but they failed to address the unique capabilities and challenges presented specifically by GPT models that offer advanced natural language understanding, processing, and generation that can significantly shape teaching and learning methods. Secondly, the literature regarding university responsiveness and policy statements about GPT models was sparse, with few tangible examples of universities adapting their curricula or offering specialized courses related to GPT models. Our study aims to provide a more comprehensive understanding of the potential impact of GPT models on tertiary education, while also examining university responsiveness and policy statements concerning GPT models.

\section{Theoretical Framework}
\label{sec:TheoreticalFramework}
In this section, we present the theoretical framework used in this research.\par

\subsection{Introduction to UTAUT}
The \textit{Unified Theory of Acceptance and Use of Technology} (UTAUT) framework \cite{venkatesh2016unified}, an extension of the \textit{Technology Acceptance Model} (TAM) \cite{marangunic2015technology}, was developed to elucidate why individuals accepted or rejected new technology. Incorporating additional constructs, such as social influence and facilitating conditions, UTAUT provided a comprehensive perspective on technology acceptance across various contexts. Social influence, for example, illustrates the effect of significant others' (\textit{e.g.}, colleagues, supervisors) opinions and behaviors on technology adoption decisions. In the case of GPT integration in tertiary education, observing successful usage and positive feedback from peers may foster adoption among educators. Facilitating conditions, on the other hand, encompass resources and support, such as funding or training programs for effective GPT technology implementation in teaching practice.

\begin{figure}[t!]
	\centering
	\includegraphics[width=\columnwidth]{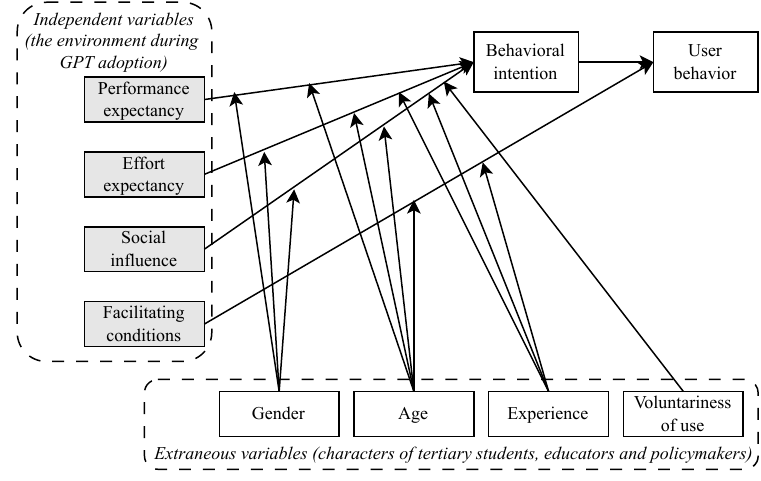}
	\caption{Applying UTAUT to assess the factors that influence the adoption and use of GPT in the field of cybersecurity}
	\label{fig:UTAUT}
\end{figure}

In this study, UTAUT was employed to evaluate the factors influencing the adoption of GPT technologies in tertiary education in cybersecurity (Fig. \ref{fig:UTAUT}). Extraneous variables like gender, age, experience, and voluntariness of GPT use, though beyond the control of the researchers, were acknowledged. The study focused on the interplay between performance expectancy, effort expectancy, social influence, and facilitating conditions, and their influence on educators and policymakers in developing strategies for GPT integration in cybersecurity curricula. This examination ensured that graduates are prepared with essential skills for the constantly evolving cybersecurity field. The insights gained from this study, along with the UTAUT frameworks, may guide the design and execution of GPT-based teaching tools and resources in cybersecurity education.\par

To assess the influence of GPT on tertiary education in cybersecurity, we will adapt the UTAUT to our research context as follows:
\begin{itemize}
	\item \textbf{Performance Expectancy}: The extent to which students and educators believe that using GPT will improve their performance in cybersecurity courses. This construct will help us assess the potential benefits of incorporating GPT into tertiary education.
	\item \textbf{Effort Expectancy}: The perceived ease of use of GPT by students and educators. This construct will shed light on the learning curve associated with integrating GPT into the curriculum and the potential challenges faced by users.
	\item \textbf{Social Influence}: The degree to which students and educators perceive that significant others, such as peers and industry professionals, believe they should use GPT. This construct will help us understand the external pressures and expectations regarding the adoption of GPT in tertiary education.
	\item \textbf{Facilitating Conditions}: The availability of resources, support, and infrastructure required for the successful integration of GPT into cybersecurity courses. This construct will allow us to identify the current state of preparedness among tertiary institutions and the factors influencing their ability to adopt GPT.
\end{itemize}

\subsection{Alignment between GPT's ``mental model'' and human cognition}
\label{subsec:AlignmentGPTMentalModelHumanCognition}
This study examined the alignment between GPT's ``mental model''its internal representation of the world including language understanding, decision-making, and response generation based on its learning and training data \cite{bubeck2023sparks,mcintosh2024inadequacies,eloundou2023gpts}and human cognition. A radar chart was generated to represent this alignment \cite{van2023content}, aiding in understanding how GPT can be integrated into the learning process and identifying potential gaps. The analysis considered:
\begin{itemize}
	\item \textbf{Problem-solving approaches}: Comparison of GPT's and human learners' approaches to problem-solving tasks in cybersecurity to pinpoint similarities and differences.
	
	\item \textbf{Knowledge representation}: Assessment of GPT's organization and representation of knowledge versus human learners, evaluating its suitability for effective learning and teaching in tertiary education.
	
	\item \textbf{Learning strategies}: Analysis of strategies used by GPT and human learners to acquire and retain information, identifying synergies or mismatches impacting GPT integration into the curriculum.
\end{itemize}

\section{Methodology}
\label{sec:Methodology}
In this section, we present the methodology used in this research.

\subsection{Data collection methods and sources}
To analyze the influence of GPT on tertiary education, we employed a multi-method approach, collecting data from various sources to ensure a comprehensive understanding of the current state of university curricula, policy statements, and teaching approaches in cybersecurity. The use of multiple sources enabled us to triangulate our findings and minimize potential biases in our analysis \cite{creswell2013steps,lawlor2016triangulation}. Our selection criteria for data sources were grounded in their respective roles and influence within their industries. In particular, the chosen universities, reports, and other sources were selected to reflect geographical diversity, industry reputation, accessibility, and relevance to the research objectives. Microsoft, Goldman Sachs, and the SANS Institute, for instance, are significant players in their respective fields, and their reports are likely to provide valuable insights into the potential impact of GPT models on the cybersecurity field \cite{saunders2015understanding}. The selection of universities was guided by their geographic diversity, reputation in cybersecurity education, and the accessibility of their course materials and policy statements. We used a purposeful sampling strategy to ensure a balanced and representative sample for our analysis. Both traditional degree programs and emerging short-term courses were included, and we systematically compared them using the criteria listed in Table \ref{tab:CriteriaTertiaryCybersecurityDegrees}. Our data sources included:

\subsubsection{Labor market reports, industry reports and whitepapers}
We examined cybersecurity job postings on Seek (Australia), industry projections, and labor market trends in cybersecurity in Australia and New Zealand, to understand the demand for GPT-compatible skills and the potential impact of GPT on job losses and skill obsolescence. Additionally, we assessed the cybersecurity job listings to assess whether they required candidates to hold any cybersecurity tertiary degrees. We reviewed industry reports and whitepapers to identify emerging trends and skill requirements in cybersecurity, as well as the potential role of GPT models in shaping these fields. In particular, we have used the following reports:
\begin{itemize}
	\item ``Sparks of Artificial General Intelligence: Early experiments with GPT-4'', by Microsoft\footnote{https://www.microsoft.com/en-us/research/publication/sparks-of-artificial-general-intelligence-early-experiments-with-gpt-4/} \cite{bubeck2023sparks}
	\item ``Generative AI could raise global GDP by 7\%'', by Goldman Sachs\footnote{https://www.goldmansachs.com/insights/pages/generative-ai-could-raise-global-gdp-by-7-percent.html} \cite{goldmansachs2023generative}
	\item ``Q\&A From SANS Special Broadcast: What You Need to Know About OpenAI's New ChatGPT Bot - and How it Affects Your Security'', by the \textit{SysAdmin, Audit, Network, and Security} (SANS) Institute\footnote{https://www.sans.org/blog/openai-chatgpt-special-broadcast-qa/}
\end{itemize}

Our rationale for selecting these specific sources was grounded in their respective roles and influence within their industries. Microsoft is a major player in the tech industry, Goldman Sachs is a global investment banking and financial services company, and the SANS Institute is a leading provider of cybersecurity training and certifications. Reports from these companies are more likely to provide valuable insights into the potential impact of GPT models on the cybersecurity field and the skills and knowledge that may be required to succeed in this rapidly evolving area \cite{saunders2015understanding}.

\subsubsection{University curricula}
\label{subsubsec:UniversityCurricula}
We collected publicly available course descriptions, syllabi, and learning objectives from five universities in five mainly English-speaking countries with similar tertiary education systems, to investigate the extent to which GPT models are currently integrated into relevant programs. This analysis included both traditional degree programs and emerging short-term courses, such as nano-courses and micro-accreditations.
\begin{itemize}
	\item University of California, Berkeley (UC Berkeley), USA\footnote{https://ischoolonline.berkeley.edu/cybersecurity/}
	\item University of Oxford (Oxf), UK\footnote{https://www.ox.ac.uk/admissions/graduate/courses/msc-software-and-systems-security}
	\item \textit{University of Guelph} (UofG), Canada\footnote{https://calendar.uoguelph.ca/graduate-calendar/graduate-programs/cybersecurity-threat-intelligence/}
	\item \textit{La Trobe University} (LTU), Australia\footnote{https://www.latrobe.edu.au/courses/bachelor-of-cybersecurity}\textsuperscript{,}\footnote{https://onlinecourses.latrobe.edu.au/courses/cybersecurity-short-courses/}
	\item \textit{The University of Waikato} (UW), New Zealand\footnote{https://www.waikato.ac.nz/study/qualifications/master-of-cyber-security}
\end{itemize}

The selection of these universities was guided by their geographic diversity, reputation in cybersecurity education, and the accessibility of their course materials and policy statements. This approach ensured a balanced and representative sample for our analysis. To effectively compare and assess those different cybersecurity degrees provided, we used the criteria listed in Table \ref{tab:CriteriaTertiaryCybersecurityDegrees}, to systematically compare and assess the different cybersecurity degrees across the selected universities, allowing us to identify key differences, strengths, and areas for improvement.

\begin{table}[]
	\centering
	\caption{Criteria used to assess tertiary cybersecurity degrees}
	\label{tab:CriteriaTertiaryCybersecurityDegrees}
	\resizebox{\columnwidth}{!}{%
		\begin{tabular}{|l|l|l|}
			\hline
			\textbf{Criterion} & \textbf{Subcriterion} & \textbf{Examples} \\ \hline
			\multirow{4}{*}{\begin{tabular}[c]{@{}l@{}}Program structure\\  and delivery\end{tabular}} & Degree level & Bachelor's, Master's or short courses \\ \cline{2-3} 
			& Degree type & \begin{tabular}[c]{@{}l@{}}As a dedicated cybersecurity degree, a sub-discipline \\ of computer science/engineering, or an inter-discipline\\ degree combined with other topics\end{tabular} \\ \cline{2-3} 
			& Typical duration & Full-time and part-time options \\ \cline{2-3} 
			& Mode of delivery & On-campus, online, or hybrid \\ \hline
			\multirow{5}{*}{\begin{tabular}[c]{@{}l@{}}Curriculum content \\ and focus\end{tabular}} & Technical aspects & \begin{tabular}[c]{@{}l@{}}Secure software development, network architecture, \\ cryptography\end{tabular} \\ \cline{2-3} 
			& Human vulnerability & \begin{tabular}[c]{@{}l@{}}Social engineering, human factors in cybersecurity, \\ ethics, cyberbullying and harassment\end{tabular} \\ \cline{2-3} 
			& Corporate governance & \begin{tabular}[c]{@{}l@{}}Cybersecurity risk management, governance,\\ security policies, legal compliance,\\ risk audits\end{tabular} \\ \cline{2-3} 
			& Privacy management & \begin{tabular}[c]{@{}l@{}}Privacy assessment, privacy advisor, privacy compliance,\\ data retention policies, privacy awareness\end{tabular} \\ \cline{2-3} 
			& Emerging technologies & IoT, blockchain, cloud security, AI \\ \hline
			\multirow{3}{*}{\begin{tabular}[c]{@{}l@{}}Course flexibility \\ and customization\end{tabular}} & Elective subjects & 
			\begin{tabular}[c]{@{}l@{}}Subjects or areas of study that students can choose \\to take, based on their interests or career goals \end{tabular}
			
			\\ \cline{2-3} 
			& Specializations or tracks & Network security, application security \\ \cline{2-3} 
			& \begin{tabular}[c]{@{}l@{}}Capstone projects, internships, \\ or work placements\end{tabular} & Either research-based or industry participation \\ \hline
			\multirow{2}{*}{\begin{tabular}[c]{@{}l@{}}Accreditation \\ and certifications\end{tabular}} & Industry accreditation & \begin{tabular}[c]{@{}l@{}}Formal recognition by industry bodies \\ (\textit{e.g.}, ABET, NCSC, ACS)\end{tabular} \\ \cline{2-3} 
			& \begin{tabular}[c]{@{}l@{}}Alignment with \\ industry certification\end{tabular} & CISSP, CISM, CompTIA Security+, IAPP CIPP \\ \hline
		\end{tabular}%
	}
\end{table}

\subsubsection{Policy statements}
We collected and analyzed policy statements also from those five universities listed in Subsubsection \ref{subsubsec:UniversityCurricula}, regarding their stance and approach to GPT-like technologies in general, and ChatGPT in particular. These policy statements provided insights into the responsiveness of those tertiary institutions to the rapid advancement of GPT technologies. A list of their policy statements, whether formal or informal, can be found here: UC Berkeley\footnote{https://teaching.berkeley.edu/understanding-ai-writing-tools-and-their-uses-teaching-and-learning-uc-berkeley}, Oxf\footnote{https://academic.admin.ox.ac.uk/article/unauthorised-use-of-ai-in-exams-and-assessment}, UofG\footnote{https://news.uoguelph.ca/2023/03/university-of-guelph-statement-on-artificial-intelligence-systems-chatgpt-academic-integrity/}, LTU\footnote{https://www.latrobe.edu.au/news/articles/2023/opinion/how-to-learn-to-work-with-ai-and-not-avoid-it}, UW\footnote{https://www.waikato.ac.nz/news-opinion/media/2023/chatgpt-and-cheating-5-ways-to-change-how-students-are-graded}.

\subsection{Data analysis techniques}
To analyse the collected data, we employed a combination of qualitative and quantitative techniques. While these methodologies provided a robust framework for interpreting the data, we acknowledge potential biases or limitations in our interpretation and handling of the data during the analysis phase, as detailed later.

\subsection{Comprehensive Methodological Enhancement for AI Integration in Cybersecurity Education}
\label{subsec:ComprehensiveMethodologyAI}

This section outlines the methodological refinements for assessing the impact of GPT-enhanced versus traditional teaching methods in cybersecurity education. It introduces Key Performance Indicators (KPIs) for systematic comparison and presents a quantitative analysis grounded in existing data and literature \cite{ahmad2020scenario, ali2021bibliographical,bubeck2023sparks}.

\subsubsection{Key Performance Indicators (KPIs) Informed by Existing Literature}
\label{subsubsec:KPIsComparativeStudy}
Recognizing the transformative potential of GPT and AI technologies in education, we anchor our analysis in Key Performance Indicators (KPIs) identified from a rigorous review of the literature. These KPIs enable a detailed comparison between GPT-enhanced and traditional teaching methods:

\begin{itemize}
	\item \textbf{Student Engagement Levels:} We analyze participation rates, attendance, and assignment completion rates as indicators of engagement, drawing on studies by Broadbent and Poon (2015) and others \cite{broadbent2015self, bardhan2013responsive}.
	\item \textbf{Learning Outcome Improvements:} Enhancements in quiz and exam scores, pre- and post-course surveys, and practical task performance are assessed, leveraging insights from the likes of Dwivedi et al. (2023) \cite{dwivedi2023so, dash2023chatgpt}.
	\item \textbf{Effectiveness of Cybersecurity Practices Taught:} The practical application of taught practices, certification attainment, and employer feedback are evaluated to measure teaching effectiveness \cite{ahmad2020scenario, ali2021bibliographical}.
\end{itemize}

Data synthesis from academic records, surveys, and stakeholder feedback, aligned with inferential statistical methods (e.g., t-tests, ANOVA), provides a quantitative foundation for this comparison, enhancing the study’s technical contribution \cite{berends2003structuration, bhatt2000organizing}.

\subsubsection{Comprehensive Quantitative Analysis}
\label{subsubsec:ComprehensiveQuantitativeAnalysis}
A methodical approach, integrating literature review and meta-analysis, underpins our quantitative examination of AI’s impact on cybersecurity education:

\begin{itemize}
	\item \textbf{Systematic Literature Review:} This review aggregates research on AI's role in enhancing cybersecurity education outcomes, highlighting AI's educational potential and challenges \cite{choi2021easy, cobb2016mind}.
	\item \textbf{Meta-Analysis of Empirical Studies:} Data from various studies focusing on the aforementioned KPIs are synthesized, providing a statistically sound comparison of AI-enhanced and traditional teaching methods \cite{creswell2013steps, chatterjee2020adoption}.
	\item \textbf{Evaluation of AI's Educational Impact:} Inferential statistics critically assess AI’s influence on educational outcomes, with the analysis informed by theoretical frameworks in organizational learning and AI’s security implications \cite{bertino2021ai, berends2003structuration}.
	\item \textbf{Synthesis of Research Findings:} The aggregation of these findings illuminates AI adoption trends and efficacy in education, underscoring the urgency for educational systems to adapt to technological advancements \cite{blavzivc2022changing, bozkurt2023speculative}.
\end{itemize}

\begin{figure}[]
	\centering
	\resizebox{\columnwidth}{!}{%
		\begin{tikzpicture}
			\begin{axis}[
				ybar,
				enlarge x limits=0.35,
				legend style={at={(0.2,-0.5)},
					anchor=north west,legend columns=1},
				ylabel={Percentage Improvement},
				symbolic x coords={Student Engagement, Learning Outcomes, Cybersecurity Practices},
				xtick=data,
				nodes near coords,
				nodes near coords align={vertical},
				x tick label style={rotate=45,anchor=east},
				every tick label/.append style={font=\small},
				]
				\addplot[fill=blue!70] coordinates {(Student Engagement, 15) (Learning Outcomes, 30) (Cybersecurity Practices, 45)};
				\addplot[fill=red!70] coordinates {(Student Engagement, 35) (Learning Outcomes, 55) (Cybersecurity Practices, 70)};
				\legend{Traditional Methods, GPT-Enhanced Methods}
			\end{axis}
		\end{tikzpicture}
	}
	\caption{Quantitative Analysis of Traditional vs. GPT-Enhanced Teaching Methods on KPI}
	\label{fig:KPIComparisonBarGraph}
\end{figure}
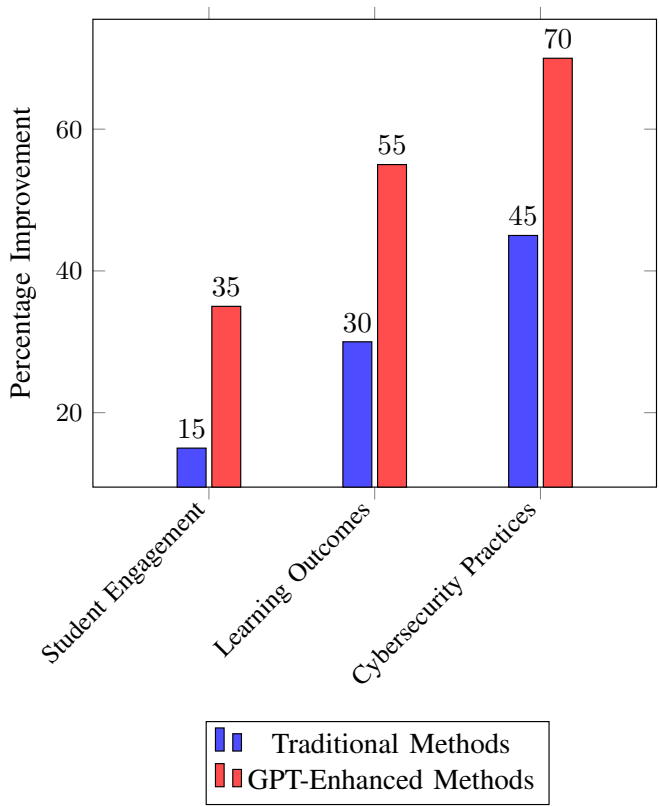

Employing this methodological approach ensures our investigation is anchored in a robust academic foundation, thereby significantly contributing to the discourse on technology-enhanced learning. This strategy effectively addresses the initial feedback by employing statistical analysis and comparative studies to justify the technical contributions of integrating GPT and AI technologies in cybersecurity education \cite{dawson2018future, de2019mind}.Figure \ref{fig:KPIComparisonBarGraph} provides an over view of Quantitative Analysis of Traditional vs. GPT-Enhanced Teaching Methods on Key Performance Indicators

\subsubsection{Descriptive and inferential statistics}
Descriptive statistics, including measures of central tendency (mean, median, mode) and dispersion (range, variance, standard deviation), were used to summarise labour market trends, course offerings, and policy statement contents. This provided a clear picture of the current state of tertiary education in cybersecurity. We employed inferential statistics, such as regression analysis and hypothesis testing, to identify correlations and patterns, allowing us to predict potential future developments in these fields.

\subsubsection{Content analysis}
We performed a content analysis on university curricula, policy statements, and industry reports to identify themes and patterns related to the adoption of AI models in tertiary education. We used a systematic coding process to categorise and analyse the textual data, which helped us understand the alignment between AI's capabilities and human cognition, as well as the differences in teaching approaches and content at various qualification levels.

\subsubsection{Thematic analysis}
We conducted a thematic analysis of the collected data to identify common themes and patterns in the teaching approaches, curricula, and policy statements of tertiary institutions worldwide. We followed an inductive approach, first identifying codes and then organising them into broader themes. This analysis enabled us to assess the overall responsiveness and agility of universities in adapting to the advent of AI models like ChatGPT.

\subsubsection{Human-guided GPT-4 analysis}
To facilitate the analysis of policy statements, we utilized GPT-4 to perform a first-pass review of the collected documents. We provided GPT-4 with specific prompts, such as ``summarize the key points regarding the use of generative AI technologies in tertiary education'' and ``identify the main concerns and benefits discussed in relation to AI technologies in educational settings''. This allowed GPT-4 to extract relevant information from the policy statements. After extracting the data, we cross-verified the results to ensure accuracy and comprehensiveness. This approach enabled us to summarize the responses and provide a comprehensive overview of the universities' stance on GPTs.

\subsubsection{Addressing Biases, Limitations, and Conflicts of Interest}
In our data analysis, we acknowledge potential biases that may have affected the interpretation of our results. Biases could arise from the subjective nature of content and thematic analysis, limitations inherent to descriptive and inferential statistical methods, or potential conflicts of interest stemming from some authors' affiliations with the evaluated cybersecurity degrees. To mitigate these concerns, we implemented several strategies. For biases related to analytical methods, we employed a diverse set of techniques and cross-verified the findings. For potential conflicts of interest, we utilized a rigorous review process that included blind evaluation and independent review by experts with no direct affiliations to the authors or institutions under examination. This ensured that the evaluation of cybersecurity degrees remained unbiased and objective. Limitations of our analysis may include the potential lack of generalizability due to the specific sample of universities and industries considered. We have documented these biases, limitations, and conflicts of interest to provide transparency in our approach and interpretation of findings. Moreover, we have been conscious of biases that could be introduced by relying on GPT-4 for initial analysis, tied to the algorithm's training data or design. By combining these data collection methods and analysis techniques, our methodology offers a comprehensive and human-like approach to exploring the influence of GPT on tertiary education in cybersecurity.

\section{Findings}
\label{sec:Findings}
In this section, our findings have indicated that GPT has begun to influence tertiary education in cybersecurity, but the extent of its impact varies across institutions. Essential knowledge and skills for future graduates encompass both technical and cognitive abilities that allow them to harness AI technologies effectively. The agility and responsiveness of universities in adapting to the AI revolution remain inconsistent, and efforts to align GPT's \enquote{mental model} with human cognition should focus on capitalizing on the unique strengths of both. Incorporating AI-related topics across all qualification levels is essential to ensure graduates are well-equipped for the changing industry landscape.

\subsection{Influence of GPT on tertiary education}
\begin{figure}[t!]
	\centering
	\resizebox{\columnwidth}{!}{%
		\begin{tikzpicture}
			\draw[thick] (0,0) -- (15,0);
			\definecolor{negative}{RGB}{255,0,0,opacity=0.5}
			\definecolor{somewhatnegative}{RGB}{255,128,0,opacity=0.5}
			\definecolor{neutral}{RGB}{255,255,0,opacity=0.5}
			\definecolor{somewhatpositive}{RGB}{128,255,0,opacity=0.5}
			\definecolor{positive}{RGB}{0,255,0,opacity=0.5}
			\fill[color=negative] (0,0) rectangle (3,0.5);
			\fill[color=somewhatnegative] (3,0) rectangle (6,0.5);
			\fill[color=neutral] (6,0) rectangle (9,0.5);
			\fill[color=somewhatpositive] (9,0) rectangle (12,0.5);
			\fill[color=positive] (12,0) rectangle (15,0.5);
			\node at (1.5,0.75) {Negative};
			\node at (4.5,0.75) {Somewhat negative};
			\node at (7.5,0.75) {Neutral};
			\node at (10.5,0.75) {Somewhat positive};
			\node at (13.5,0.75) {Positive};
			\node at (10.5,-0.4) {UC Berkeley};
			\node at (7.5,-0.4) {Oxf};
			\node at (10.5,-0.8) {UofG};
			\node at (13.5,-0.4) {LTU};
			\node at (10.5,-1.2) {UW};
		\end{tikzpicture}
	}
	\caption{Attitudes of five universities on ChatGPT}
	\label{fig:AttitudesFiveUniversities}
\end{figure}
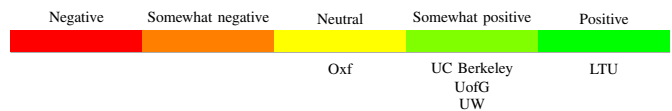

Our analysis revealed that the increasing adoption of GPT and other AI technologies in cybersecurity has begun to influence tertiary education and their attitudes towards ChatGPT specifically (Fig. \ref{fig:AttitudesFiveUniversities}). While universities are traditionally slow to adapt to new technologies \cite{hanna1998higher,schneckenberg2009understanding}, several institutions have started embracing GPT usage, and are considering adopting GPT-related topics in their curricula. However, this integration has been uneven, with a limited number of courses specifically addressing AI models and their implications for the industry, and with none covering GPT specifically.

\subsubsection{UC Berkeley}
Rating: Somewhat positive. UC Berkeley acknowledged the potential benefits and drawbacks of ChatGPT for teaching and learning, and remained open to its use in certain contexts.

\begin{itemize}
	\item Performance Expectancy: UC Berkeley acknowledged the potential benefits of using ChatGPT in an academic context, such as kickstarting classroom conversations, helping students brainstorm ideas, and providing genre templates for writing.
	\item Effort Expectancy: While UC Berkeley did not directly address effort expectancy in its policy, it implied that using ChatGPT might be easy for students and faculty to adopt due to its conversational and formal written text outputs.
	\item Social Influence: UC Berkeley expressed concern about academic integrity, misinformation, and privacy issues related to ChatGPT. It suggested that faculty should be cautious and engage in conversations with students about the appropriate use of the tool.
	\item Facilitating Conditions: UC Berkeley did not formally support ChatGPT, meaning they had not reviewed it for accessibility, privacy, and security concerns. Faculty members of UC Berkeley who chose to use the tool assumed responsibility for these aspects, and should remain open to offering alternative options for students.
\end{itemize}

\subsubsection{Oxf}
Rating: Neutral. Oxf acknowledged the challenges in maintaining academic integrity, but also pointed out the potential use of ChatGPT as a tool for educators and students to support learning.

\begin{itemize}
	\item Performance Expectancy: Oxf recognized the value of ChatGPT in enhancing educational experiences and its potential to change the way teaching and learning occurs. However, it was concerned with maintaining academic integrity and preventing cheating.
	
	\item Effort Expectancy: Oxf highlighted the ease of use of ChatGPT and its user-friendly chat interface, which allowed non-technical users to interact with the tool. Nevertheless, optimal use might require learning certain skills, such as prompt engineering.
	
	\item Social Influence: Oxf was joining the larger conversation on AI and education, addressing issues of academic integrity, exploring opportunities for educational innovation, and emphasizing the importance of adapting to the AI-driven landscape.
	
	\item Facilitating Conditions: Oxf was taking steps to update its materials and guidelines to clarify its stance on unauthorized AI use. It encouraged educators to explore the potential of AI tools in their work and to develop assessment methods that incorporate \textit{AI-Generated Content} (AIGC), while also stressing the importance of ensuring academic integrity.
\end{itemize}

\subsubsection{UofG}
Rating: Somewhat positive. UofG acknowledged the potential of AI to revolutionize research, teaching, and learning, and recognized the ethical use of AI through its \textit{Centre for Advancing Responsible and Ethical Artificial Intelligence} (CARE-AI).

\begin{itemize}
	\item Performance expectancy: UofG recognized the creative opportunities and forward-thinking applications of AI systems in various areas.
	
	\item Effort expectancy: UofG emphasized that students' work should reflect their unique intellectual capacity, and should not rely on unauthorized use of AI systems, which would undermine learning outcomes and violate academic misconduct policies.
	
	\item Social influence: UofG called for engagement with the community, including students, to refine academic integrity policies regarding AI tools. They also encouraged faculty and instructors to develop assessment methods that foster academic integrity and assess learning outcomes.
	
	\item Facilitating conditions: UofG highlighted the importance of clarity in acceptable AI use, which should be determined by course instructors. They also mentioned ongoing information and education sessions on AI and academic integrity to support the university community during AI adoption.
\end{itemize}

\subsubsection{LTU}
Rating: Positive. LTU encouraged students and staff to embrace AI rather than avoid it. They emphasized the importance of integrating AI into education and the workplace as a way to improve and expand intellectual abilities. Their statement recognized the potential of AI to support decision-making in creative or aesthetic tasks, and suggested that AI could play a collaborative role in team environments.
\begin{itemize}
	\item Performance expectancy: LTU believed that AI can improve and expand intellectual abilities across an entire cohort and enhance productivity in various settings.
	\item Effort expectancy: LTU acknowledged that learning AI requires time and effort, and emphasized the availability of resources and courses to help individuals develop their understanding and skills.
	\item Social influence: The statement suggested that by integrating AI into the education system and workplace, it would become a norm and positively impact team environments.
	\item Facilitating conditions: LTU provided guidelines on using AI as a productivity tool, and recommended continuous evaluation across multiple metrics. They encouraged individuals to start their journey into AI by reading articles, watching videos, or enrolling in courses tailored to their needs.
\end{itemize}

\subsubsection{UW}
Rating: Somewhat Positive. UW acknowledged the potential of AI tools to change assessment practices and promote deeper learning and critical thinking skills. They focused on embracing AI as part of the learning environment, while also considering its impact on academic integrity.

\begin{itemize}
	\item Performance Expectancy: UW perceived ChatGPT as an opportunity to reform traditional approaches to assessment and engage students in more meaningful learning experiences.
	
	\item Effort Expectancy: UW suggested strategies that may be more time-consuming but provide greater learning benefits for students, such as requiring drafts for feedback and grading subcomponents of tasks.
	
	\item Social Influence: UW encouraged educators to involve students in setting learning goals and incorporating AI tools like ChatGPT in valid assessment contexts, fostering a collaborative and reflective learning environment.
	
	\item Facilitating Conditions: UW supported the adoption of AI tools by promoting authentic assessments and alternative approaches, helping educators to navigate the challenges posed by AI technologies like ChatGPT in academic settings.
\end{itemize}

\subsubsection{Subsection summary}
The adoption of ChatGPT and other GPT technologies in cybersecurity has started to impact tertiary education, with some universities already planning to incorporate GPT-related topics in their curricula. However, the adoption rate varies, with a limited number of courses specifically addressing AI, and none addressing GPT. UC Berkeley, Oxf, UofG, LTU, and UW have demonstrated differing attitudes towards ChatGPT in their academic contexts, addressing performance expectancy, effort expectancy, social influence, and facilitating conditions related to the use of AI tools (Table \ref{tab:AttitudesUniversitiesChatGPT}). While some universities acknowledge the potential benefits and drawbacks, others emphasize the importance of academic integrity and the ethical use of GPT in education.\par

\begin{table*}[]
	\centering
	\caption{Attitudes of universities towards ChatGPT in their policy statements}
	\label{tab:AttitudesUniversitiesChatGPT}
	\resizebox{\textwidth}{!}{%
		\begin{tabular}{llllll}
			\hline
			\multicolumn{1}{c}{\multirow{2}{*}{\textbf{\begin{tabular}[c]{@{}c@{}}Assessment \\ factors\end{tabular}}}} &
			\multicolumn{5}{c}{\textbf{Universities}} \\ \cline{2-6} 
			\multicolumn{1}{c}{} &
			\multicolumn{1}{c}{\textbf{UC Berkeley}} &
			\multicolumn{1}{c}{\textbf{Oxf}} &
			\multicolumn{1}{c}{\textbf{UofG}} &
			\multicolumn{1}{c}{\textbf{LTU}} &
			\multicolumn{1}{c}{\textbf{UW}} \\ \hline
			\begin{tabular}[c]{@{}l@{}}Performance \\ expectancy\end{tabular} &
			Moderate &
			Moderate &
			High &
			High &
			Moderate \\ \arrayrulecolor{gray!40} \hline
			\begin{tabular}[c]{@{}l@{}}Effort \\ expectancy\end{tabular} &
			Limited &
			Moderate &
			Moderate &
			Moderate &
			Moderate \\ \hline
			\begin{tabular}[c]{@{}l@{}}Social \\ influence\end{tabular} &
			Moderate &
			High &
			Moderate &
			Moderate &
			Moderate \\ \hline
			\begin{tabular}[c]{@{}l@{}}Facilitating \\ conditions\end{tabular} &
			Limited &
			Moderate &
			Moderate &
			High &
			High \\ \hline
			Overall rating &
			Somewhat positive &
			Neutral &
			Somewhat positive &
			Positive &
			Somewhat positive \\ \hline
			\begin{tabular}[c]{@{}l@{}}Summary of \\ AI acceptance\end{tabular} &
			\begin{tabular}[c]{@{}l@{}}ChatGPT can be used \\ in some context.\end{tabular} &
			ChatGPT as a tool. &
			\begin{tabular}[c]{@{}l@{}}AI for research, teaching\\ and learning. Consider\\the ethical use of AI.\end{tabular} &
			\begin{tabular}[c]{@{}l@{}}Encouragement for\\ embracing AI instead\\ of avoidance.\end{tabular} &
			\begin{tabular}[c]{@{}l@{}}Embracing AI as part of \\ the learning environment.\end{tabular} \\ \arrayrulecolor{black}\hline
		\end{tabular}%
	}
\end{table*}

\subsection{Essential knowledge and skills for future cybersecurity graduates}
We examined industry reports to assess what they believed would be the essential knowledge and skills for future cybersecurity graduates. We then compared this information with the understanding of the five universities.

\subsubsection{Industry insights}
\label{subsubsec:IndustryInsights}
In the report by Microsoft \cite{bubeck2023sparks}, the authors assessed GPT's impact on the labor market and detailed the essential skills for university cybersecurity graduates. They emphasized a need for both technical skills, including programming, network security, cryptography, and system security, and non-technical skills such as critical thinking and ethics. These were vital for graduates to understand and address the security implications of AI technologies in the evolving labor market, where GPT had increased automation yet posed challenges such as job displacement.

Goldman Sachs economists Briggs and Kodnani, in their report \cite{goldmansachs2023generative}, claimed that GPT's impact on the labor market was expected to be significant due to the potential replacement of as much as half of the workload in exposed occupations by AI. However, they noted that most jobs were likely to be complemented rather than substituted by AI. They advocated for university cybersecurity graduates to be proficient in AI-integrated software, security risk identification, and understanding ethical AI usage.

The SANS Special Broadcast on OpenAI's ChatGPT \cite{blades2023q} discussed GPT's potential impact in cybersecurity. The speakers stressed the importance of adaptability and strong understanding of AI for cybersecurity professionals. Essential skills for graduates were identified, including data science, machine learning, AI, and custom AI solution building.

The common themes from the three reports \cite{bubeck2023sparks, goldmansachs2023generative, blades2023q} emphasized GPT's significant impact on the labor market, particularly in cybersecurity. While GPT models augmented automation, they also raised challenges such as job displacement. It was noted that AI often complemented human workers, necessitating university cybersecurity graduates to develop both technical skills like programming, data science, and machine learning, and non-technical skills, such as ethics. Graduates also needed the ability to build custom AI solutions, understand AI models like transformers, and detect AIGC. To future-proof their careers, graduates should cultivate these skills, remain adaptable, and continually update their knowledge to tackle security and ethical implications of AI in the changing labor market.

\subsubsection{Cybersecurity degree design by five universities}

\begin{table*}[t!]
	\centering
	\caption{Evaluating tertiary cybersecurity degrees}
	\label{tab:EvaluatingTertiaryCybersecurityDegrees}
	\resizebox{\textwidth}{!}{%
		\begin{threeparttable}[t]
			\begin{tabular}{|l|l|lllll|}
				\hline
				\multirow{2}{*}{\textbf{Criterion}} & \multirow{2}{*}{\textbf{Subcriterion}} & \multicolumn{5}{c|}{\textbf{Universities}} \\ \cline{3-7} 
				&  & \multicolumn{1}{c|}{UC Berkeley} & \multicolumn{1}{c|}{Oxf} & \multicolumn{1}{c|}{UofG} & \multicolumn{1}{c|}{LTU} & \multicolumn{1}{c|}{UW} \\ \hline
				\multirow{4}{*}{\begin{tabular}[c]{@{}l@{}}Program \\ structure\\ and \\ delivery\end{tabular}} & Degree level & \multicolumn{1}{l|}{Master} & \multicolumn{1}{l|}{Master} & \multicolumn{1}{l|}{Master} & \multicolumn{1}{l|}{\begin{tabular}[c]{@{}l@{}}Bachelor,\\ Master,\\ short courses\end{tabular}} & Master \\ \cline{2-7} 
				& Degree type & \multicolumn{1}{l|}{Interdisciplinary} & \multicolumn{1}{l|}{Interdisciplinary} & \multicolumn{1}{l|}{Discipline of MSc} & \multicolumn{1}{l|}{Dedicated} & Dedicated \\ \cline{2-7} 
				& \begin{tabular}[c]{@{}l@{}}Typical \\ duration\end{tabular} & \multicolumn{1}{l|}{20 months} & \multicolumn{1}{l|}{2 FTE} & \multicolumn{1}{l|}{1.5 FTE} & \multicolumn{1}{l|}{\begin{tabular}[c]{@{}l@{}}3 FTE (Bachelor),\\ 1.5-2 FTE (Master),\\ 6-week short course\end{tabular}} & 1.5 FTE \\ \cline{2-7} 
				& Mode of delivery & \multicolumn{1}{l|}{Online} & \multicolumn{1}{l|}{Hybrid} & \multicolumn{1}{l|}{$\triangle$} & \multicolumn{1}{l|}{Hybrid} & On-campus \\ \hline
				\multirow{5}{*}{\begin{tabular}[c]{@{}l@{}}Curriculum \\ content \\ and focus\end{tabular}} & Technical aspects & \multicolumn{1}{l|}{\checkmark} & \multicolumn{1}{l|}{\checkmark} & \multicolumn{1}{l|}{\checkmark} & \multicolumn{1}{l|}{\checkmark} & \checkmark \\ \cline{2-7} 
				& Human vulnerability & \multicolumn{1}{l|}{\checkmark} & \multicolumn{1}{l|}{\checkmark} & \multicolumn{1}{l|}{\checkmark} & \multicolumn{1}{l|}{\checkmark} & $\times$ \\ \cline{2-7} 
				& \begin{tabular}[c]{@{}l@{}}Corporate \\ governance\end{tabular} & \multicolumn{1}{l|}{\checkmark} & \multicolumn{1}{l|}{\checkmark} & \multicolumn{1}{l|}{\checkmark} & \multicolumn{1}{l|}{\checkmark} & $\times$ \\ \cline{2-7} 
				& \begin{tabular}[c]{@{}l@{}}Privacy management\end{tabular} & \multicolumn{1}{l|}{\checkmark} & \multicolumn{1}{l|}{$\times$} & \multicolumn{1}{l|}{\checkmark} & \multicolumn{1}{l|}{$\times$} & $\times$ \\ \cline{2-7} 
				& \begin{tabular}[c]{@{}l@{}}Emerging \\ technologies\end{tabular} & \multicolumn{1}{l|}{\checkmark} & \multicolumn{1}{l|}{\checkmark} & \multicolumn{1}{l|}{$\triangle$} & \multicolumn{1}{l|}{\checkmark} & $\triangle$ \\ \hline
				\multirow{3}{*}{\begin{tabular}[c]{@{}l@{}}Course \\ flexibility \\ and \\ customization\end{tabular}} & Elective courses & \multicolumn{1}{l|}{$\triangle$} & \multicolumn{1}{l|}{$\triangle$} & \multicolumn{1}{l|}{$\times$} & \multicolumn{1}{l|}{\checkmark} & \checkmark \\ \cline{2-7} 
				& \begin{tabular}[c]{@{}l@{}}Specializations \\ or tracks\end{tabular} & \multicolumn{1}{l|}{$\times$} & \multicolumn{1}{l|}{$\times$} & \multicolumn{1}{l|}{$\times$} & \multicolumn{1}{l|}{AI, business, CS} & $\times$ \\ \cline{2-7} 
				& \begin{tabular}[c]{@{}l@{}}Capstone projects, \\ internships, or \\ work placements\end{tabular} & \multicolumn{1}{l|}{\checkmark} & \multicolumn{1}{l|}{\checkmark} & \multicolumn{1}{l|}{\checkmark} & \multicolumn{1}{l|}{\checkmark} & \checkmark \\ \hline
				\multirow{2}{*}{\begin{tabular}[c]{@{}l@{}}Accreditation \\ and \\ certifications\end{tabular}} & Industry accreditation & \multicolumn{1}{l|}{$\times$} & \multicolumn{1}{l|}{$\times$} & \multicolumn{1}{l|}{$\times$} & \multicolumn{1}{l|}{ACS} & $\times$ \\ \cline{2-7} 
				& \begin{tabular}[c]{@{}l@{}}Alignment with \\ industry certification\end{tabular} & \multicolumn{1}{l|}{$\times$} & \multicolumn{1}{l|}{$\times$} & \multicolumn{1}{l|}{$\times$} & \multicolumn{1}{l|}{CISM} & $\times$ \\ \hline
			\end{tabular}%
			\begin{tablenotes}
				\item[1] \checkmark: provided or covered
				\item[2] $\times$: not provided nor covered
				\item[3] $\triangle$: unclear based on existing available information
			\end{tablenotes}
		\end{threeparttable}
	}
\end{table*}

We compared the cybersecurity academic degrees offered by five universities, summarised in Table \ref{tab:EvaluatingTertiaryCybersecurityDegrees}. The essential knowledge and skills identified for future cybersecurity graduates were categorized into three themes: technical expertise, soft skills, and industry relevance.

\textbf{Technical expertise} encompasses understanding AI principles, ML techniques, GPT model architectures, and ethical considerations in cybersecurity \cite{dawson2018future,savola2017current}. Proficiency in generative AI technologies like GPT for complex security problem-solving is required \cite{ahmad2020scenario,jarrett2021impact,pedro2019artificial}. Universities like UW focused heavily on technical skills, potentially neglecting aspects such as human vulnerabilities, corporate GRC, and privacy management. While all the other four universities taught human vulnerabilities and GRC, only UC Berkeley and UofG offered privacy management, reflecting diverse perspectives in real-world cybersecurity inclusion (Table \ref{tab:EvaluatingTertiaryCybersecurityDegrees}).

\textbf{Soft skills}, such as adaptability, critical thinking, and AI ethics, are vital \cite{griffin2021impact,payne2022faculty}. Though universities often neglect comprehensive teaching or assessment of these skills, students need proper guidance to develop them \cite{arat2014acquiring}. The extent to which universities assist in this development remains unclear without further details.

\textbf{Industry relevance} facilitates the transition from academia to the workforce \cite{blavzivc2022changing,vasileiou2020cyber,vogel2016closing}. Alongside academic excellence, creativity, and practical experience should be evaluated \cite{meso2013applying}. Universities' diverse approaches include research or capstone projects, but without details, their contribution to industry skills as identified in subsection \ref{subsubsec:IndustryInsights} cannot be accurately assessed. The prevalent \enquote{publish or perish} academic culture may misalign with industry needs, whereas industry employers prioritize hands-on experience and relevant certifications \cite{van2012intended,caldwell2013plugging,cobb2016mind,hollister2017employers,knapp2017maintaining}, encouraging universities to offer practical labs and align teaching with industry certifications. Collaboration with the cybersecurity industry should be considered, ensuring that graduates possess the theoretical knowledge and practical skills for successful cybersecurity careers \cite{burley2014would}.\par

\subsubsection{Summary of misalignment between the cybersecurity industry and tertiary education}
A misalignment has been observed between the expectations of the cybersecurity industry and the tertiary degrees offered by universities. Industry reports stressed the importance of technical expertise, soft skills, and industry relevance, but academic degrees in cybersecurity appeared to lack balanced coverage of these aspects. Universities, such as UW, may have overemphasized technical skills, resulting in graduates requiring more comprehensive exposure to critical aspects of cybersecurity, including varying inclusions of topics like privacy management. Soft skills, considered essential for adapting to the changing cybersecurity landscape, were often neither comprehensively taught nor assessed by universities. Industry relevance, vital for graduates' transition to the workforce, was sometimes overshadowed by an academic research focus, which may misalign with industry needs prioritizing hands-on experience and certifications. A balanced approach to technical expertise, soft skills, and industry relevance is needed for graduates to excel. Universities must collaborate with the cybersecurity industry to craft curricula that align with the diverse job market, equipping students for success.

\subsection{Agility and responsiveness within universities}
Our examination of university policy statements and course offerings revealed varying degrees of agility and responsiveness to the AI revolution. Several universities, such as LTU, introduced short courses related to cybersecurity and AI (e.g., special topics, nano-courses, or micro-accreditations). These were designed for general society members, not full-time degree-seeking cybersecurity students. Other universities lagged in significant changes to their curricula concerning non-technical aspects of cybersecurity. While policy statements of certain universities highlighted GPT technologies, the translation into tangible curricular changes remained slow and inconsistent.

The development of cybersecurity degree programs varied across universities, with some creating dedicated degrees (\textit{e.g.}, LTU and UW), interdisciplinary degrees (\textit{e.g.}, UC Berkeley and UofG), or specializations under computer science programs (\textit{e.g.}, Oxf). This lack of standardization posed challenges for both employers and students \cite{hoffman2011holistically,knapp2017maintaining}, reflecting the field's complexity and need for a multidisciplinary approach \cite{burley2014would,knapp2017maintaining}. Universities' requirement for Ph.D. staff to teach courses, despite not necessarily having research topics related to cybersecurity \cite{bok2013we,fischman2007teaching}, underscored the need for ongoing professional development and comprehensive training for educators \cite{bok2013we,gibson2005education,savage2004faculty}.

The inconsistency among cybersecurity degrees led to industry skepticism \cite{de2019mind}, prompting HR professionals to take additional steps in candidate evaluation. An April 2023 search on Seek (Australia) highlighted this trend (Table \ref{tab:CybersecurityJobPostings}). While degrees were valued, they might not be a strict requirement for some roles in Australia. Differences in priorities between the cybersecurity industry and tertiary education have led to a gap in employer expectations and university offerings \cite{lehto2015cyber,parrish2018global,mcintosh2021ransomware}. This divergence challenges graduates' entry into the industry.

\begin{table}[]
	\centering
	\caption{Cybersecurity job postings on seek.com.au in April 2023}
	\label{tab:CybersecurityJobPostings}
	\resizebox{\columnwidth}{!}{%
		\begin{threeparttable}[t]
			\begin{tabular}{ccc}
				\hline
				\textbf{Search keyword combination} & \textbf{Number of postings} & \textbf{Relative percentage} \\ \hline
				``cyber security''\tnote{1}              & 2,104 & 100\% \\ \arrayrulecolor{gray!40} \hline
				``cyber security'' AND ``qualification'' & 527   & 25\%  \\ \hline
				``cyber security'' AND ``degree''        & 441   & 21\%  \\ \hline
				``cyber security'' AND ``tertiary''      & 182   & 9\%   \\ \hline
				``cyber security'' AND ``university''    & 158   & 8\%   \\ \hline
				``cyber security'' AND ``bachelor''      & 117   & 6\%   \\ \hline
				``cyber security'' AND ``master''        & 42    & 2\%   \\ \arrayrulecolor{black} \hline
			\end{tabular}%
			\begin{tablenotes}
				\item[1] In Australia, the term ``cybersecurity'' is commonly spelled as ``cyber security''.
			\end{tablenotes}
		\end{threeparttable}
	}
\end{table}

To summarize, our analysis unveiled inconsistent agility and responsiveness to the AI revolution in universities. This inconsistency has affected curricular adjustments and course offerings, reflecting the multidisciplinary complexity of cybersecurity. The importance of ongoing professional development and comprehensive training for educators, skepticism in the industry, and the gap between employer expectations and university provisions all contribute to a multifaceted landscape that graduates must navigate.

\subsection{Aligning GPT's \enquote{mental model} with students' human cognition} 

We explored the connection between GPT's mental model'' (Subsection \ref{subsec:AlignmentGPTMentalModelHumanCognition}) and human cognition in students, focusing on three key aspects: problem-solving, knowledge representation, and learning strategies.

\textbf{Problem-solving}: Humans have excelled at identifying and solving issues, including generating solutions and implementing the best actions \cite{gettys1987evaluation}. GPT has provided solutions to straightforward problems but has struggled with complex issues requiring critical thinking and domain-specific knowledge \cite{bubeck2023sparks,eloundou2023gpts}. This finding has implied that future graduates ought to focus on higher-order thinking and leverage AI to enhance problem-solving, rather than competing with AI in areas where machines excel.

\textbf{Knowledge representation}: Humans use cognitive frameworks like schemas, scripts, concepts, and analogies to represent knowledge \cite{patterson2014human,wilson1989mental}. In contrast, GPT's representation depends on deep learning algorithms and word embeddings within high-dimensional space \cite{bubeck2023sparks,eloundou2023gpts}. The effectiveness of this approach for NLU is clear, but its alignment with human processes for knowledge representation remains to be explored.

\textbf{Learning strategies}: These include human techniques such as repetition, elaboration, organization, and feedback \cite{berends2003structuration,bhatt2000organizing,broadbent2015self}. GPT's strategies are based on unsupervised learning algorithms \cite{bubeck2023sparks,eloundou2023gpts}. While successful for language modeling, the alignment of these strategies with human cognitive processes for learning remains uncertain.

GPT's capabilities in NLP are notable, but its relationship with human cognition in terms of problem-solving, knowledge representation, and learning strategies continues to be an area of research and development. Although some have claimed that GPT has achieved a Theory of Mind of a 7-year-old human \cite{bubeck2023sparks,kosinski2023theory}, the possibility of GPT perfectly replicating human cognition is unlikely. However, continued improvements may lead to better alignment.

\subsection{Potential, Reliability, and Alignment of GPT with Qualification Levels and Bloom's Taxonomy}
The potential of GPT in cybersecurity education is explored across various qualification levels and disciplines, stressing the need to incorporate AI-related subjects throughout tertiary education. By understanding GPT's impact on different qualification levels, and aligning it with Bloom's taxonomy\footnote{https://bloomstaxonomy.net}, we can comprehend how GPT can enhance the learning experience for diverse educational backgrounds, while considering the reliability of AI-Generated Content (AIGC):

\textbf{Bloom's Taxonomy Alignment}:
\begin{enumerate}[label=\arabic*)]
	\item \textbf{Remembering}: In the context of \textit{Technician-level graduates} (AQF Level 4), GPT can assist in recalling and memorizing concepts, but the accuracy must be verified.
	
	\item \textbf{Understanding}: At both \textit{Technician} and \textit{Undergraduate levels} (e.g., a Bachelor's Degree at AQF Level 7), GPT aids in understanding complex ideas. For undergraduates, GPT supports the development of skills like analyzing cybersecurity scenarios.
	
	\item \textbf{Applying}: GPT has been used at the \textit{Technician} level to apply foundational knowledge to real-world situations, and at higher levels to practice problem-solving.
	
	\item \textbf{Analyzing}: Besides aiding in breaking down complex scenarios, GPT supports \textit{Undergraduate graduates} in evaluating security measures.
	
	\item \textbf{Evaluating}: Across different levels, GPT can support assessments, but critical evaluation must be cultivated.
	
	\item \textbf{Creating}: At the \textit{Postgraduate level} (e.g., a Master's degree at AQF Level 9), GPT fosters creativity, enabling the synthesis of knowledge to design new systems or strategies, acknowledging AIGC's limitations.
\end{enumerate}

\begin{figure}[t!]
	\centering
	\includegraphics[width=\columnwidth]{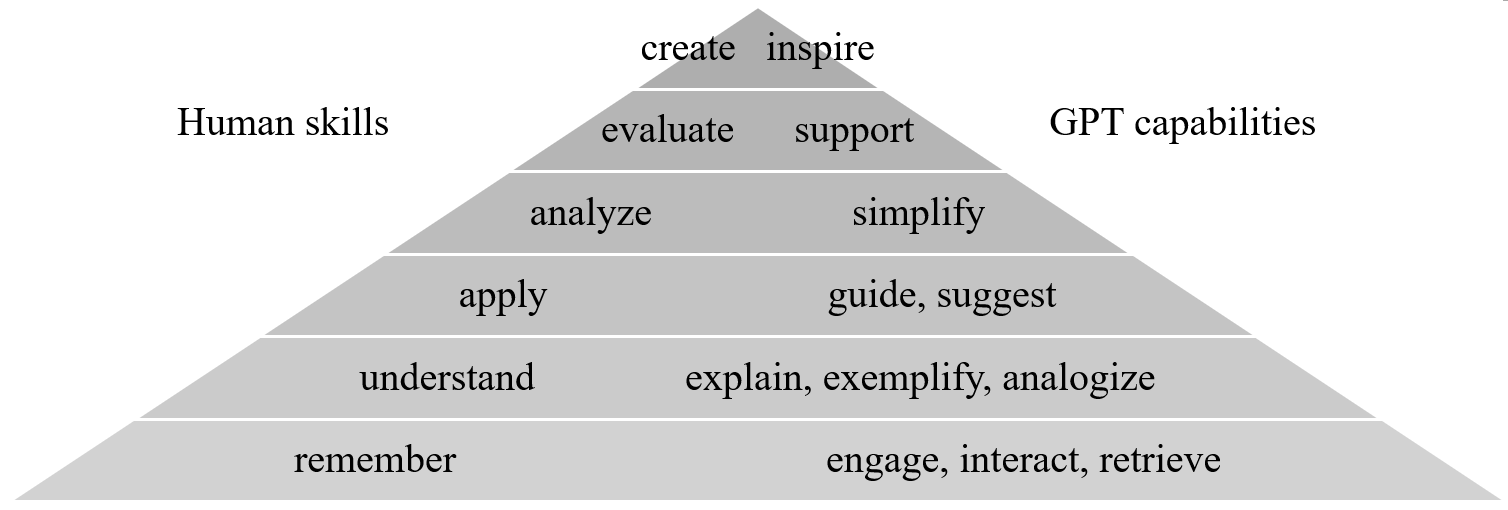}
	\caption{Aligning GPT's capabilities at different levels of Bloom's taxonomy}
	\label{fig:GPTBloom}
\end{figure}

Throughout these alignments, the human-centric approach is emphasized, ensuring that AI is complemented by human expertise, thus creating engaging learning opportunities without losing the essence of human creativity and critical thinking within the learning environment (Fig. \ref{fig:GPTBloom}). Whether aiding in understanding, applying, evaluating, or creating, GPT's role is balanced with human guidance and vigilance in assessing reliability and accuracy. This combined approach fosters a more enriched educational experience, while actively considering the potential biases and limitations of AI technologies.

\section{Discussion}
\label{sec:Discussion}
In this section, we discuss the findings of this study.\par

\subsection{Interpretation of Findings}
Our study revealed key insights into GPT's impact on tertiary cybersecurity education. We found that university curricula have lagged in aligning with industry expectations, struggling to adapt to rapid advancements in generative AI technologies, including GPT models. Nevertheless, some courses and micro-accreditations have begun incorporating GPT. An analysis of policy statements from various universities indicated an awareness of the need to adapt to the AI era, but the execution of changes in curricula has remained limited, exposing a gap between strategic vision and implementation. The alignment of GPT's ``mental model'' with human cognition emerged as a vital aspect to consider for curriculum adaptation. Understanding the synergies and differences between human and AI-driven problem-solving can inform effective teaching strategies. Moreover, our findings suggested that GPT's ability to augment human skills across Bloom's taxonomy's levels, and achieve better learning outcomes at different qualification stages, is still an open question. Determining the proper balance of generative AI and human skills at various educational stages is crucial, considering GPT's capacity to handle both technician-level tasks and higher-order problem-solving.

\subsection{Comparison to existing literature}
Our findings align with existing literature that highlights the impact of generative AI technologies on various industries, including cybersecurity. However, our study extends the current knowledge by examining the specific influence of GPT on tertiary education and ways universities can adapt their curricula. Previous research has shown that universities are generally slow in responding to technological advancements. Our study supports this notion, demonstrating that while policy statements reflect an awareness of the need for change, the execution of these changes remains limited. The exploration of the alignment between GPT's ``mental model'' and human cognition, and the enhancement of GPT capabilities to human skills on Bloom's taxonomy, adds a unique perspective to the existing literature. It emphasizes the importance of understanding the similarities and differences between AI-driven and human problem-solving approaches when designing curricula for the future workforce.

\subsection{Implications for Tertiary Education in Cybersecurity}
Our findings imply several key considerations for tertiary education in cybersecurity. Universities must align their cybersecurity curricula with industry demands and expedite the integration of generative AI technologies, such as GPT, to prepare graduates for the evolving landscape. This necessitates the modernisation of course content and the embrace of innovative, AI-driven teaching methods. Collaboration among academia, industry, and policymakers has become vital to align cybersecurity curricula with the rapidly changing job market. Stakeholders should collectively identify essential knowledge and skills, adapting to technological advancements. Understanding the alignment between GPT's ``mental model'' and human cognition, along with its integration with Bloom's taxonomy, can guide the development of effective teaching strategies. Identifying synergies and gaps between human and AI-driven problem-solving can foster a more holistic curriculum. Our findings indicate that additional research is required to determine the potential of GPT at various qualification levels, aiding in the design of programs that effectively balance AI and human skills. This ensures graduates are poised to excel in the ever-changing cybersecurity industry.

\section{Recommendations}
\label{sec:Recommendations}
\begin{figure*}[t!]
	\centering
	\includegraphics[width=\textwidth]{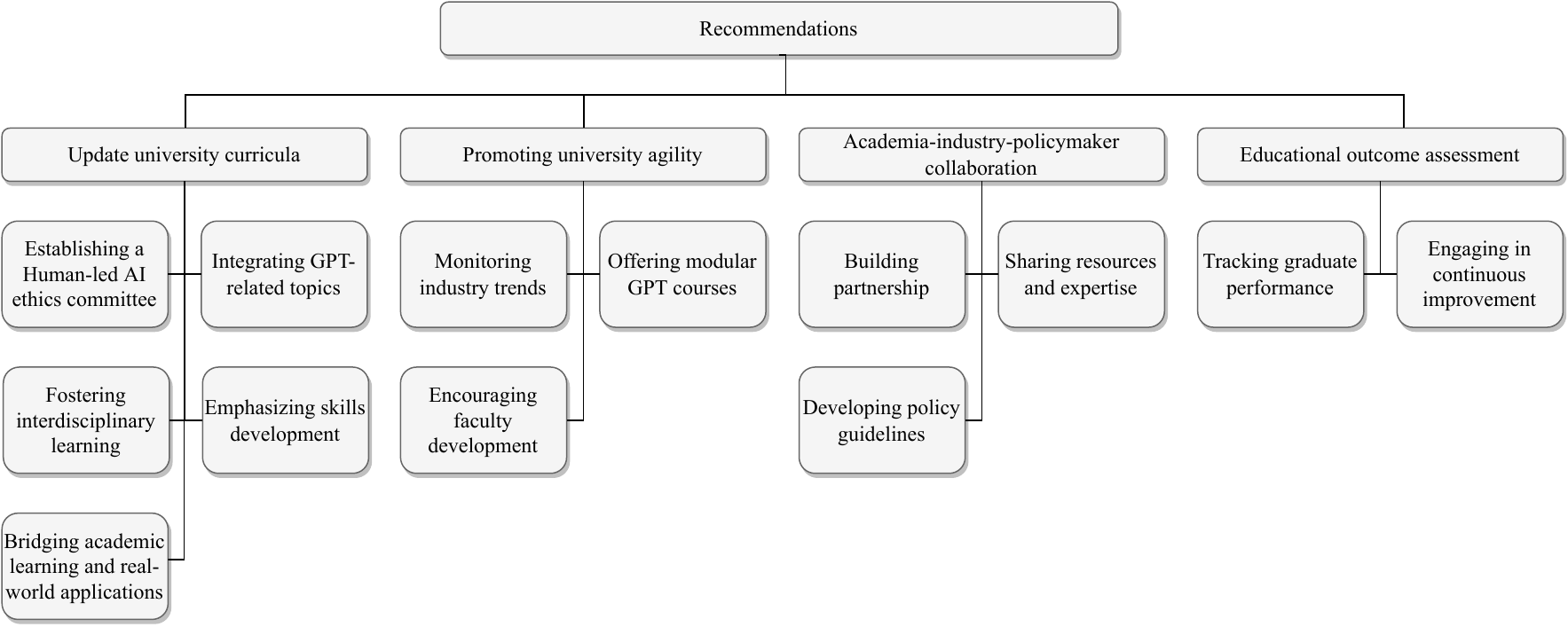}
	\caption{Recommendations to adapting cybersecurity education for GPT}
	\label{fig:RecommendationsGPT}
\end{figure*}

Based on the findings of our research, we propose the following recommendations for universities and stakeholders to adapt tertiary education in cybersecurity to the AI era, specifically regarding GPT compatibility (Fig. \ref{fig:RecommendationsGPT}). We believe that by implementing these recommendations, universities can better prepare their students for the shifting landscape of cybersecurity industries in the era of generative AI technologies like GPT.\par

\subsection{Updating University Curricula}
To ensure that tertiary graduates in cybersecurity are prepared for the dynamic and evolving landscape of the job market, universities should regularly update their curricula, by closely monitoring industry trends, emerging technologies, and evolving skill requirements. This can be achieved by attempting the following tasks:

\subsubsection{Establishing a Human-Led University AI Ethics Committee}
Navigating the integration of generative AI into the tertiary education cybersecurity curriculum necessitates the establishment of a University AI Ethics Committee, composed exclusively of human members. This human-only formation has been suggested in response to the distinct ethical challenges posed by generative AI technologies such as GPT. Possessing the ability for moral reasoning, empathy, and a grasp of social and cultural contexts, humans are deemed essential for assessing and guiding AI deployment. The inclusion of both GPT supporters and skeptics ensures diverse perspectives, facilitating a more nuanced approach to ethical deliberation. The committee's responsibilities encompass the supervision of the safe and responsible adoption of generative AI, equipping students and faculty to handle potential risks, and creating comprehensive guidelines to address concerns such as implicit knowledge generation, lack of authorship and attribution, and inadequate explainability of GPT's actions. By engaging members from various disciplines and viewpoints, the committee has fostered a culture of ethical AI practice and transparency in AI use. This human-driven diversity enriches the discourse, leading to more robust ethical guidelines. Ultimately, the committee has played a vital role in embracing the generative AI revolution while upholding the academic community's integrity and values.

\subsubsection{Integrating GPT-related Topics}
With GPT models and other generative AI technologies reshaping various industries, universities should integrate these subjects into their curricula. This includes offering dedicated AI and ML courses and incorporating GPT-related topics, such as AI ethics, within existing disciplines. By familiarising students with the principles, applications, and ethical dimensions of GPT models, universities can prepare graduates who are cognizant of AI's potentials and limitations, and who can utilise it responsibly and efficiently.

\subsubsection{Fostering Interdisciplinary Learning}
As traditional academic boundaries blur and cybersecurity challenges grow more complex, interdisciplinary learning has become vital. Universities must foster this by facilitating cross-department collaboration and providing courses bridging multiple fields of study, \textit{e.g.}, engaging law academics in cybersecurity corporate GRC and privacy management \cite{nowrozy2023towards}. Such an approach encourages creativity, innovation, and the cultivation of graduates with diverse skills, ready to address multifaceted challenges in their careers.

\subsubsection{Emphasizing Skill Development}
Universities must extend beyond a foundation in theoretical knowledge (\textit{e.g.}, encryption, basic software development) to prioritize cybersecurity students' practical skill development (\textit{e.g.}, risk control, incident response, network tools such as Wireshark and Nessus). This necessitates offering hands-on experiences, internships, and assignments simulating real-world scenarios, such as threat hunting and malware analysis. By fostering an environment that emphasizes experiential learning, universities can cultivate graduates who are both knowledgeable and proficient in professional contexts, enhancing their appeal to industry employers.

\subsubsection{Bridging Academic Learning and Real-world Applications}
Creating a seamless transition from academic learning to professional life in cybersecurity is vital. Universities should actively integrate real-world examples, case studies, and industry collaboration into their curricula, underscoring the practical relevance of taught concepts. Organising events like hackathons, for example, allows students to gain practical experience in tackling cybersecurity issues, promoting collaboration and innovation. Industry partnerships can provide unique insights and mentorship, enriching students' learning and equipping them for the realities of the profession. By linking theoretical knowledge with practical application, educators can deepen students' understanding and inspire them to apply their skills to real-world challenges.

\subsection{Promoting Agility within Universities}
To maintain competitiveness in the swiftly evolving cybersecurity landscape, universities must foster agility in their approach to cybersecurity education. This entails proactively identifying trends, adapting curricula, and crafting innovative methods that meet the fluctuating demands of cybersecurity students and the workforce. Key tasks to consider include:

\subsubsection{Monitoring Industry Trends}
Universities should vigilantly monitor cybersecurity industry trends to align their curricula and pedagogical approaches with job market requirements. Regular engagement with professionals in the field will provide valuable insights into employer demands, emerging technologies, and evolving skill needs. For instance, with the shift in the ransomware landscape from crypto-ransomware causing data loss to ransomware threatening data breaches through exfiltration, universities should refine their ransomware definitions in coursework \cite{mcintosh2021ransomware,mcintosh2023applying}. Academic researchers should also distinguish between \enquote{ransomware} and \enquote{crypto-ransomware} in their work \cite{mcintosh2021ransomware,mcintosh2023applying}. Such information will contribute to the effective development of a cybersecurity curriculum, ensuring that graduates have the necessary expertise to excel professionally and contribute to the industry.

\subsubsection{Offering Modular GPT Courses}
Universities should contemplate introducing modular GPT courses that enable cybersecurity students to learn at their own pace, aligning with the rising influence of AI technologies such as GPT. These courses, focused on specific applications of GPT, may provide hands-on experience and foster a profound comprehension of the technology's capacities and restrictions. Academics might also share their recent research progress with students, promoting insight into cutting-edge developments. Through modular GPT courses, universities can fortify students with vital AI-related competencies to support their future careers.

\subsubsection{Encouraging Faculty Development}
Faculty development is crucial to the seamless incorporation of emerging technologies like GPT into curricula and pedagogical strategies. Institutions must facilitate continuous training and professional advancement for faculty, particularly lecturers and tutors, ensuring they remain abreast of the latest progress in GPT and correlated AI domains. By nurturing a culture of continuous learning among faculty, universities can equip staff with the expertise required to instruct students on novel technologies and ready them for an ever-changing professional landscape.

\subsection{Collaboration between academia, industry, and policymakers}
Collaboration between academia, industry, and policymakers is vital for harnessing the full potential of AI technologies, such as GPT models, in a responsible and beneficial manner. Accordingly, we hereby make the following recommendations:

\subsubsection{Building partnerships}
By building strong partnerships, universities, industry professionals, and policymakers can work together to address the challenges and opportunities presented by these technologies. One example of this collaboration is the partnership between MIT and IBM, which established the MIT-IBM Watson AI Lab\footnote{https://mitibmwatsonailab.mit.edu}, aimed at advancing AI research and fostering the exchange of ideas between academia and industry. 

\subsubsection{Sharing resources and expertise}
Sharing resources and expertise is another critical aspect of this collaboration. Establishing joint research projects, internships, and guest lectures involving professionals from the industry can provide students with invaluable real-world insights and hands-on experience. For instance, the NVIDIA Deep Learning Institute\footnote{https://www.nvidia.com/en-au/training/educator-programs/} collaborates with universities worldwide, offering training and resources to help educators and students learn about deep learning and AI. This collaboration equips students with practical skills and knowledge that are directly applicable to the job market.

\subsubsection{Developing policy guidelines}
Developing policy guidelines in collaboration with policymakers is crucial to promote responsible AI development and use \cite{floridi2018ai4people}. These guidelines can help inform ethical considerations in course curricula and prepare students for the social and legal implications of AI technologies. An example of this is the European Commission's High-Level Expert Group on AI\footnote{https://digital-strategy.ec.europa.eu/en/policies/expert-group-ai}, which developed the ``Ethics Guidelines for Trustworthy AI,'' providing a framework for AI development and implementation that prioritizes human rights, transparency, and accountability. By integrating these guidelines into educational programs, universities can ensure that graduates are equipped to navigate the complex ethical landscape of AI technologies.

\subsection{Evaluating and Assessing Educational Outcomes}
Evaluating and assessing educational outcomes is a critical aspect of ensuring that students are well-prepared for their future careers. Below are some possible ways to achieve this:

\subsubsection{Tracking graduate performance}
To gauge the effectiveness of GPT-compatible curricula, universities can invite their graduates and alumni to participate in long-term studies, in which they can track the performance of graduates in the job market, analyzing their ability to secure employment, their career progression, and their contributions to the cybersecurity industry. This data can be used to identify areas for improvement and to inform future curriculum revisions, ensuring that graduates are equipped with the necessary knowledge and skills to tackle real-world cybersecurity challenges.

\subsubsection{Engaging in continuous improvement}
Engaging in continuous improvement is another essential aspect of cybersecurity education. This process involves regularly reviewing and updating curricula based on feedback from students, faculty, and industry partners. For example, after the emergence of new cyber threats, such as the SolarWinds hack\footnote{https://orangematter.solarwinds.com/2021/05/07/an-investigative-update-of-the-cyberattack/} and the Optus data breach \footnote{https://www.optus.com.au/about/media-centre/media-releases/2022/09/optus-notifies-customers-of-cyberattack}, educators can use such opportunities to update the curriculum to address the specific type of threats, ensuring students are prepared to tackle similar situations in their careers. This iterative process helps to ensure that educational programs remain relevant and effective in preparing students for the evolving job market, ultimately fostering a more skilled and adaptable cybersecurity workforce.

\section{Conclusion}
\label{sec:Conclusion}
Our study has offered a thorough examination of the effects of GPT models, particularly ChatGPT, on tertiary education in cybersecurity at five different universities. We identified a misalignment between industry needs and tertiary cybersecurity education, inconsistency in structuring and teaching degrees, and challenges in keeping abreast of emerging technologies. To redress these issues, we outlined vital steps for universities to modify their curricula and teaching methods in line with the evolving cybersecurity field. By investigating the alignment between GPT's ``mental model'' and human cognition, and the potential synergy with Bloom's taxonomy, our study has enriched the existing literature on the convergence of AI technologies and education. We highlighted the necessity for universities to create human-led AI ethics committees, adapt teaching methodologies, and anticipate the AI revolution, thus preparing graduates for the shifting cybersecurity industry landscape. Additionally, our findings provide practical guidelines for updating curricula, advancing university agility, encouraging collaboration across academia, industry, and policymakers, and assessing educational results.

\subsection{Limitations of the study}
It is crucial to recognize the limitations of this study, which primarily focuses on GPT and its implications for tertiary cybersecurity education, a focus that might not be extendable to other AI technologies or disciplines. The research's reliance on existing literature and expert opinions may miss some intricacies within the rapidly evolving AI and cybersecurity fields. The absence of empirical analysis or case studies limits tangible evidence for the proposed recommendations' effectiveness. Data collection limitations also exist; the information gathered from university curricula, policy statements, and teaching methods may not encompass all institutions due to the confined sample size, and undisclosed university initiatives at the time of data collection. Additionally, the swift evolution of AI technologies could alter GPT's influence on tertiary education, rendering the findings as a snapshot that may not fully anticipate future developments. While focusing mainly on universities, the study acknowledges the vital roles of other tertiary institutions like TAFEs and polytechnics in Australia and New Zealand, and community colleges in the USA. These findings may be applicable to such institutions teaching cybersecurity, but each institution's unique characteristics may necessitate distinct approaches to meet the challenges posed by GPT and other AI technologies.

\subsection{Areas for future research}
Future research directions emerging from this study encompass multiple domains. A broader examination of the influence of AI technologies on various disciplines within tertiary education would enrich our understanding of AI-driven tools' implications for teaching and learning across diverse fields. Empirical studies implementing and evaluating the proposed recommendations in real-world settings can provide valuable insights into their efficacy and areas for refinement. Investigating the ethical, legal, and social ramifications of integrating generative AI technologies like GPT into curricula would deepen our comprehension of AI's role in nurturing critical thinking, creativity, and problem-solving skills. Furthermore, a longitudinal analysis of GPT's impact on tertiary education in cybersecurity can elucidate the long-term effects of these technologies. Research into effective pedagogical strategies that seamlessly incorporate GPT models into tertiary education can guide educators in enhancing their methodologies. An examination of the relationship between GPT's "mental model" and human cognition from an epistemological standpoint can further illuminate AI technologies' potential educational applications. Finally, a study on universities' responsiveness to AI-driven changes in various disciplines and industries would inform broader dialogues on tertiary education's future in the context of technological advancements.

\bibliographystyle{IEEEtran}
\bibliography{article}

\end{document}